\documentclass{sf2a-conf2021}
\usepackage[utf8]{inputenc}
\usepackage[T1]{fontenc}
\usepackage{mathrsfs}
\usepackage{amsmath}
\usepackage{amssymb}
\usepackage{graphicx}
\usepackage{tikz}
\usepackage{hyperref}
\usepackage[]{natbib}
\usepackage{epstopdf}
\usepackage{bm}

\def\BibTeX{{\rm B\kern-.05em{\sc i\kern-.025em b}\kern-.08em
    T\kern-.1667em\lower.7ex\hbox{E}\kern-.125emX}}
\bibpunct{(}{)}{;}{a}{}{,}  

\begin{document}

\TitreGlobal{SF2A 2021}


\title{Dipolar magnetic fields in binaries and gravitational waves}

\runningtitle{Dipolar magnetic fields in binaries and gravitational waves}

\author{A. Bourgoin$^{1,}$}\address{SYRTE, Observatoire de Paris, PSL Research University, CNRS, Sorbonne Universit\'es, UPMC Univ. Paris 06, LNE, 61 avenue de l'Observatoire, 75014 Paris, France}
\address{Département d'Astrophysique-AIM, CEA/DRF/IRFU, CNRS/INSU, Université Paris-Saclay, Université de Paris, 91191 Gif-sur-Yvette, France}

\author{C. Le Poncin-Lafitte$^1$}

\author{S. Mathis$^2$}

\author{M.-C. Angonin$^1$}




\setcounter{page}{237}


\maketitle


\begin{abstract}
The LISA mission will observe gravitational waves emitted from tens of thousands of galactic binaries, in particular white dwarf binary systems. These objects are known to have intense magnetic fields. However, these fields are usually not considered as their influence on the orbital and rotational motion of the binary is assumed for being too weak. It turns out that magnetic fields modify the orbits, in particular their geometry with respect to the observer. In this work, we revisit the issue, assuming magnetostatic approximation, and we show how the magnetic fields within a binary system generate a secular drift in the argument of the periastron, leading then, to modifications of the gravitational waveforms that are potentially detectable by LISA.
\end{abstract}



\section{Introduction}
  Laser Interferometer Space Antenna (LISA) is an ESA L class mission that aims at observing gravitational waves (GW) from space \citep{2017arXiv170200786A} in the frequency band from below $10^{-4}\ \mathrm{Hz}$ to above $10^{-1}\ \mathrm{Hz}$. Among the different sources of GW that LISA will observe are the galactic binaries. They comprise primarily white dwarfs (WD) but also neutron stars (NS) and stellar-origin black holes in various combinations. For LISA's frequency window, a typical orbital period for galactic binaries of WD and NS is ranging from tenth of hour to tenth of seconds corresponding to a semi-major axis distribution between $10^4\ \mathrm{km}$ to $10^{6}\ \mathrm{km}$. Therefore, LISA will observe GW emitted by galactic binaries during the inspiral phase, just before merger which is detected by ground-based GW detectors such as LIGO, Virgo, KRAGA, and the future Einstein Telescope. LISA will thus bring precious informations about the long term evolution of galactic binaries and their internal structure and equation of state.
  
  It is expected that the galaxy is populated with approximately hundred millions of WD-WD systems and millions of NS-WD binaries \citep{2009CQGra..26i4030N}. These compact objects can have intense magnetic fields that may reach up to $10^9\ \mathrm{G}$ for WD and up to $10^{15}\ \mathrm{G}$ for NS. White dwarfs with magnetic field ranging from $10^6\ \mathrm{G}$ to $10^9\ \mathrm{G}$ should represent around $20\%$ of the total WD population while NS with magnetic field between $10^{14}\ \mathrm{G}$ to $10^{15}\ \mathrm{G}$ should represent around $10\%$ of the total NS population \citep{2020AdSpR..66.1025F}. Therefore, a non-negligible amount of galactic binaries that LISA will observe can potentially be made of highly magnetized objects. Since then, the impact of the magnetic effects on the GW signal must be investigated. Indeed, the future data processing of the LISA mission will require that all observable physical effects be modeled with a sufficient accuracy in order to better understand the physics of these compact objects and to process the foreground noise coming from galactic binaries. Preliminary studies have focused on the monochromatic approximation only \citep{PhysRevD.76.083006,LDCGroup}, that is to say the well-known Keplerian motion. However, it has been shown that a number of physical effects, such as the back reaction due to gravitational radiation \citep{2009CQGra..26i4030N} or the dynamical tides \citep{2011MNRAS.412.1331F,2014MNRAS.444.3488F,2020MNRAS.491.3000M}, can also have a significant impact for the time span of the LISA mission. Therefore, isolating and deriving the influence of magnetic effects on the orbital motion in order to assess their observability through the GW spectrum, is thus of a prime importance in the context of the future data processing of the LISA mission.
  
  With this goal in mind, we explore the impact of dipolar magnetic fields on the orbital evolution of a binary system, within the magnetostatic approximation. We focus our attention on secular effects only, and we consider the case where the system has not reach equilibrium yet (i.e., that the magnetic torques acting on the magnetic moments do not cancel out). Then, the effect on the GW signal due to the magnetic perturbation is assessed.
  
\section{Orbit and spins evolution considering magnetic interaction}
\label{sec:dynamics}

\begin{figure}[t]
  \centering
  \includegraphics[scale=0.22]{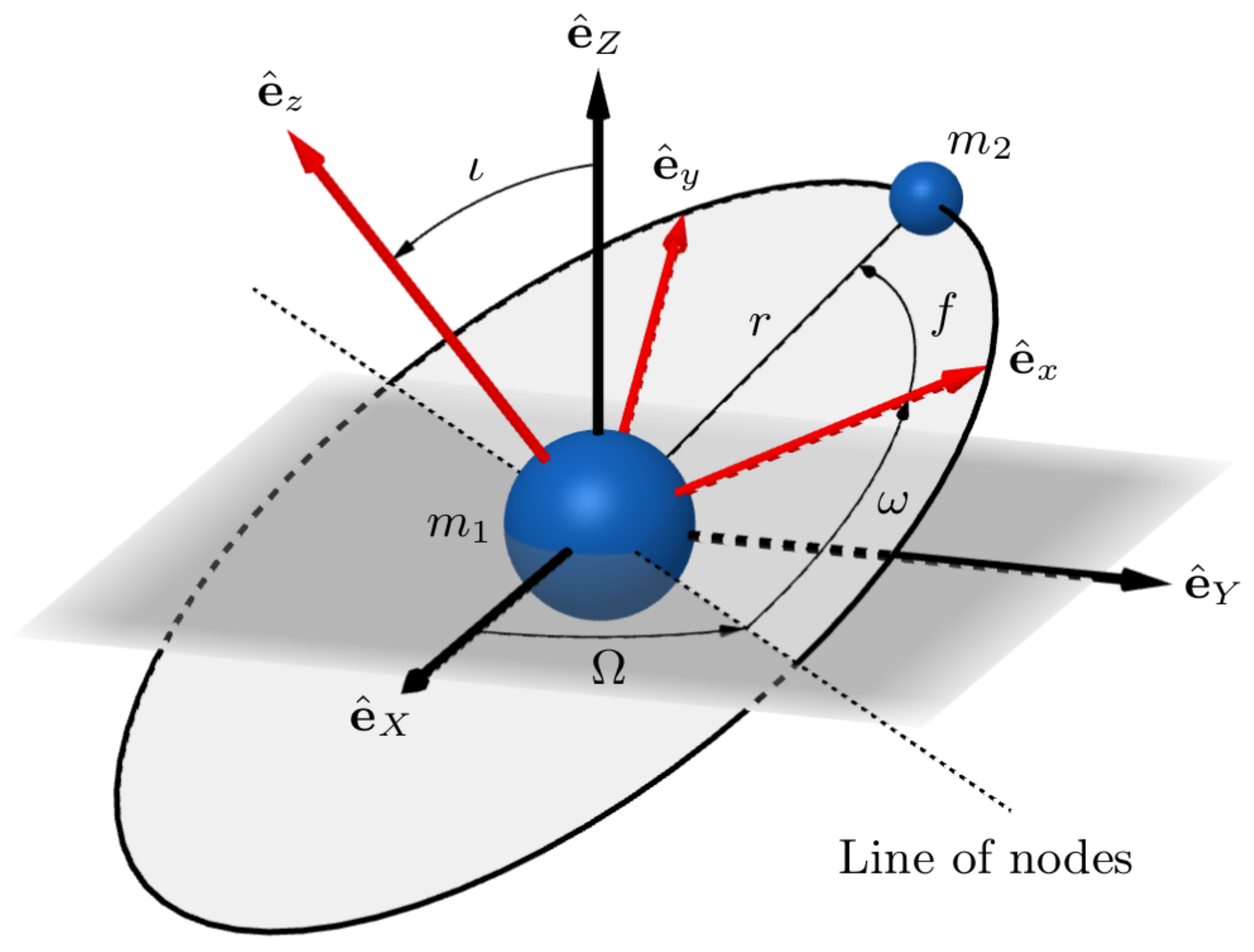}
  \includegraphics[scale=0.22]{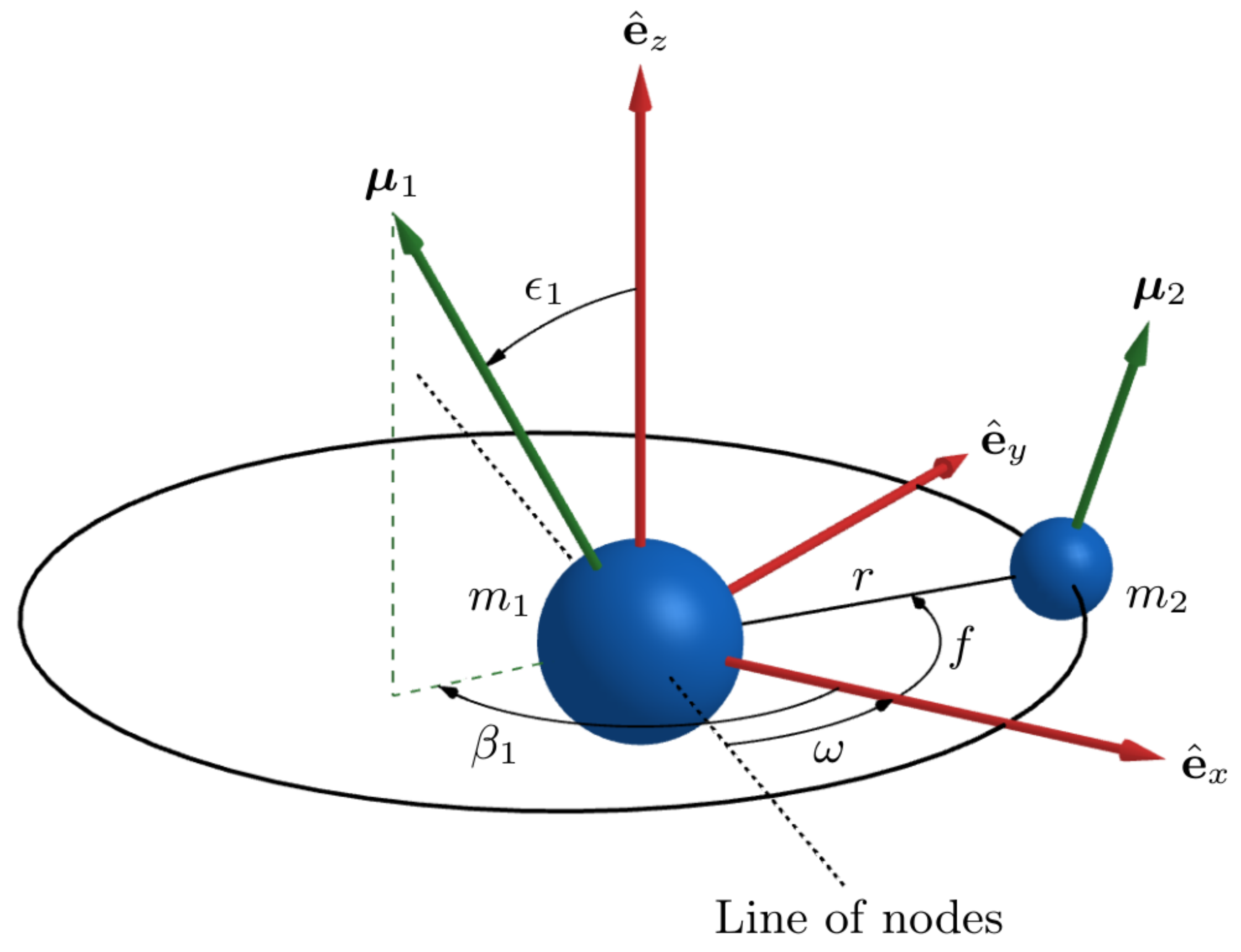}\vspace{-0.1cm}
  \caption{{\bf Left:} Orientation of the orbit frame $(\hat{\mathbf e}_x,\hat{\mathbf e}_y,\hat{\mathbf e}_z)$ in the center-of-mass frame $(\hat{\mathbf e}_X,\hat{\mathbf e}_Y,\hat{\mathbf e}_Z)$. The primary is shown at the center-of-mass of the binary system for simplicity. {\bf Right:} Orientation of the magnetic moments in the orbit frame $(\hat{\mathbf e}_x,\hat{\mathbf e}_y,\hat{\mathbf e}_z)$. The obliquity $\epsilon_1$ and the precession angle $\beta_1$ are represented for the primary only. The obliquity is a tilt between the normal to the orbital plane and the direction of the magnetic moments and the precession angle is the separation between the direction of $\hat{\mathbf e}_x$ and the projection of the magnetic moments on the orbital plane.}
  \label{fig:orbit}
\end{figure}

We consider two compact and well-separated bodies that are forming a binary system (cf., Fig. \ref{fig:orbit}). The system consists of a first body (the primary) of mass $m_1$ and dipolar magnetic moment $\bm{\mu}_1$, and a second body (the secondary) of mass $m_2$ and dipolar magnetic moment $\bm{\mu}_2$. We consider a relativistic description of the point-mass dynamics up to 2.5 post-Newtonian (2.5PN) approximation \citep{1990PhRvD..42.1123L,2014LRR....17....2B}. In that sense we treat the dynamics of the binary system coherently with the loss of energy that is radiated away by gravitational waves. We also consider magnetic fields within the magnetostatic approximation which corresponds to the case of fossil fields that are frozen into the stars (e.g., \citet{2004Natur.431..819B,2010ApJ...724L..34D}). In other words, we follow \citet{1990MNRAS.244..731K} and assume that the internal currents that generate the magnetic field of the primary are not significantly distorted by the external field of the secondary and conversely. In addition, we consider that the magnetic fields of both stars are dominated by their dipole moments $\bm{\mu}_1$ and $\bm{\mu}_2$. This is coherent with the fossil-field hypothesis. Since internal currents are assumed to be stationary, the magnitude of the magnetic moments (i.e., $\mu_1=|\bm\mu_1|$ and $\mu_1=|\bm\mu_2|$) is taken to be constant during the motion so that we focus on their orientation only. We consider for simplicity that the direction of $\bm{\mu}_1$ and $\bm{\mu}_2$ are aligned with $\bm S_1$ and $\bm S_2$ being the spins of the primary and secondary respectively. In order to follow the evolution of the magnetic moments, we introduce two angles per stars, namely the obliquities $\epsilon_1$ and $\epsilon_2$, and the precession angles $\beta_1$ and $\beta_2$ (the angles are depicted in Fig.~\ref{fig:orbit} for the primary only).

The methodology proceeds as follows. First, we adopt osculating elements within the method of variation of arbitrary constants in order to describe the orbital motion beyond the Keplerian approximation. We are thus dealing with a set of six first order differential equations describing changes of six Keplerian elements due to relativistic and magnetic terms. Obviously, the magnetic contribution depends on obliquities and precession angles of the magnetic moments. Therefore, we consider four additional differential equations describing changes in the orientation of $\bm{\mu}_1$ and $\bm{\mu}_2$. Secondly, in order to assess a long term secular description of the motions, the equations are averaged over the true anomaly (namely $f$ in Fig. \ref{fig:orbit}). This allows us to show that, on one hand, general relativity secularily affects the shape of the orbit (i.e., $a$ the semi-major axis and $e$ the eccentricity) and $\omega$ the argument of the periastron (see also \citet{1990PhRvD..42.1123L}). On the other hand, magnetism does affect the orientation of the orbit (i.e., $\iota$ the inclination, $\Omega$ the longitude of the node, and the argument of the periastron) but not the shape. Then, in order to solve for the secular evolution of the orbit, we first solve for the orientation of magnetic moments which can be mainly decoupled from the orbital motion. We show that the rate of precession of the magnetic moments are approximately given by
\begin{equation}
  \dot\beta_1\propto\frac{\mu_0}{4\pi}\frac{\mu_1\mu_2}{S_1}\frac{\cos\epsilon_{20}}{a^3(1-e^2)^{3/2}}\mathrm{,} \qquad \dot\beta_2\propto\frac{\mu_0}{4\pi}\frac{\mu_1\mu_2}{S_2}\frac{\cos\epsilon_{10}}{a^3(1-e^2)^{3/2}}\mathrm{,}
\end{equation}
where $\epsilon_{10}$ and $\epsilon_{20}$ are the initial value of the obliquities of the primary and secondary respectively, and where $\mu_0$ is the vacuum permeability. We used the notation $S_1=|\bm S_1|$ and $S_2=|\bm S_2|$.

Then, the solutions for the orientation are then substituted within the orbital equations of motion which in turn are solved at first order in the perturbations. We show that the only angle which is secularly affected by magnetism is the argument of the periastron which presents a rate of precession that is approximately given by
\begin{equation}
  \dot\omega\propto\frac{3\mu_0}{4\pi\sqrt{G}}\frac{\mu_1\mu_2}{m_1m_2}\frac{\sqrt{m_1+m_2}}{a^{7/2}(1-e^2)^2}\cos\epsilon_{10}\cos\epsilon_{20}\mathrm{,}
  \label{eq:dome}
\end{equation}
where $G$ is the gravitational constant.

Finally, after substituting the rate of precession of the argument of the periastron from Eq. \ref{eq:dome} into the expression for the well-known GW mode polarizations $h_+$ and $h_\times$, we are able to assess the observability of magnetic effects by the future LISA mission. This is the subject of the following section.

\section{Observing binary's magnetic effect in GW with LISA}

The rate of precessions that have been derived in the last section are all proportional to the product of the magnitude of the magnetic moments of the primary and secondary. A rough estimate of the amplitude of the magnetic moments is assessed with $\mu=(2\pi/\mu_0)BR^3$ (see e.g., \citet{2019MNRAS.488...64P}), where $R$ is the equatorial radius of the star and $B=|\mathbf{B}|$ is the magnitude of the magnetic field. Considering that magnetic fields can reach up to $10^9\ \mathrm{G}$ for the most magnetized WD and up to $10^{15}\ \mathrm{G}$ for the most magnetized NS, we have the following rough estimates:
\begin{equation}
  \mu_{\mathrm{WD}}\sim 10^{33}\ \mathrm{A}\cdot\mathrm{m}^2\left(\frac{R_{\mathrm{WD}}}{1\ R_{\oplus}}\right)^3\left(\frac{B_{\mathrm{WD}}}{10^9\ \mathrm{G}}\right)\mathrm{,} \qquad \mu_{\mathrm{NS}}\sim 10^{30}\ \mathrm{A}\cdot\mathrm{m}^2\left(\frac{R_{\mathrm{NS}}}{10\ \mathrm{km}}\right)^3\left(\frac{B_{\mathrm{NS}}}{10^{15}\ \mathrm{G}}\right)\mathrm{.}
  \label{eq:magmomWD}
\end{equation}
Therefore, even though the magnetic fields can be several orders of magnitude stronger for highly magnetized NS than for highly magnetized WD, the magnetic moment can be higher for WD since it evolves as the cubic power of the radius and is only linear in the magnitude of the magnetic field (see also \citet{2018ApJ...868...19W}).

We thus consider a double WD system where the mass of the primary is $m_1=1.1\ M_{\odot}$ and the mass of the secondary is $m_2=0.4\ M_{\odot}$ such that the total mass is $m=1.5\ M_{\odot}$. We consider a system with high magnetic moments at the level of $\mu_1=10^{33}\ \mathrm{A}\cdot\mathrm{m}^2$ and $\mu_2=5\times 10^{32}\ \mathrm{A}\cdot\mathrm{m}^2$. In order to probe the LISA frequency window from $10^{-4}$ to $10^{-1}\ \mathrm{Hz}$, we assume initial values of the semi-major axis ranging from $a_0=10^{6}\ \mathrm{km}$ to $a_0=10^4\ \mathrm{km}$, respectively. The dynamics of this system is modeled according to discussion of Sect. \ref{sec:dynamics}. From the dynamics, the GW mode polarizations are determined with Einstein's quadrupole formula. Finally, the relative errors generated by magnetism on $h_+$ and $h_\times$ are computed as follows:
\begin{equation}
  \mathrm{err}(h_+)=\frac{|(h_+)_{\mathrm{GR}+\mathrm{M}}-(h_+)_{\mathrm{GR}}|}{(h_+)_{\mathrm{GR}+\mathrm{M}}}\mathrm{,} \qquad \mathrm{err}(h_\times)=\frac{|(h_\times)_{\mathrm{GR}+\mathrm{M}}-(h_\times)_{\mathrm{GR}}|}{(h_\times)_{\mathrm{GR}+\mathrm{M}}}\mathrm{,}
\end{equation}
where $(h_+)_{\mathrm{GR}+\mathrm{M}}$ is the mode polarization calculated accounting for both GR and magnetic dipole-dipole interaction while $(h_+)_{\mathrm{GR}}$ contains the gravitational contribution only. The same notation is used for $h_\times$.

\begin{figure}[t]
    \centering
    \includegraphics[scale=0.91]{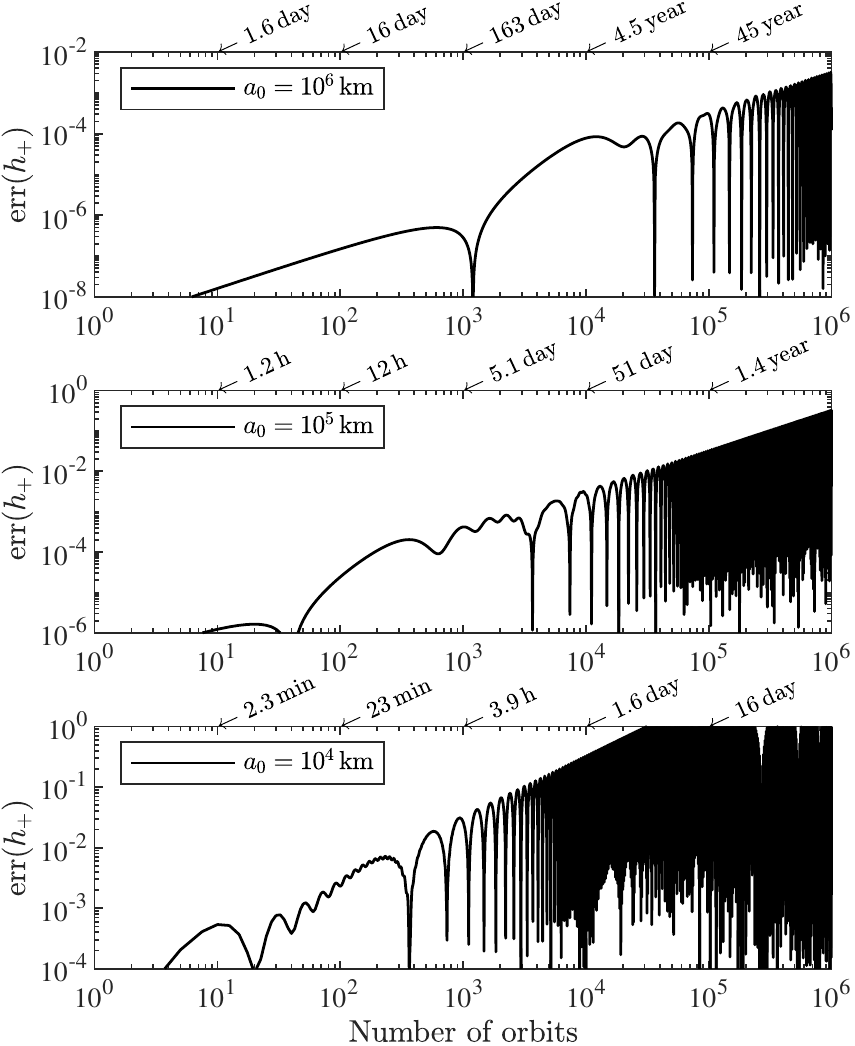}
    \includegraphics[scale=0.91]{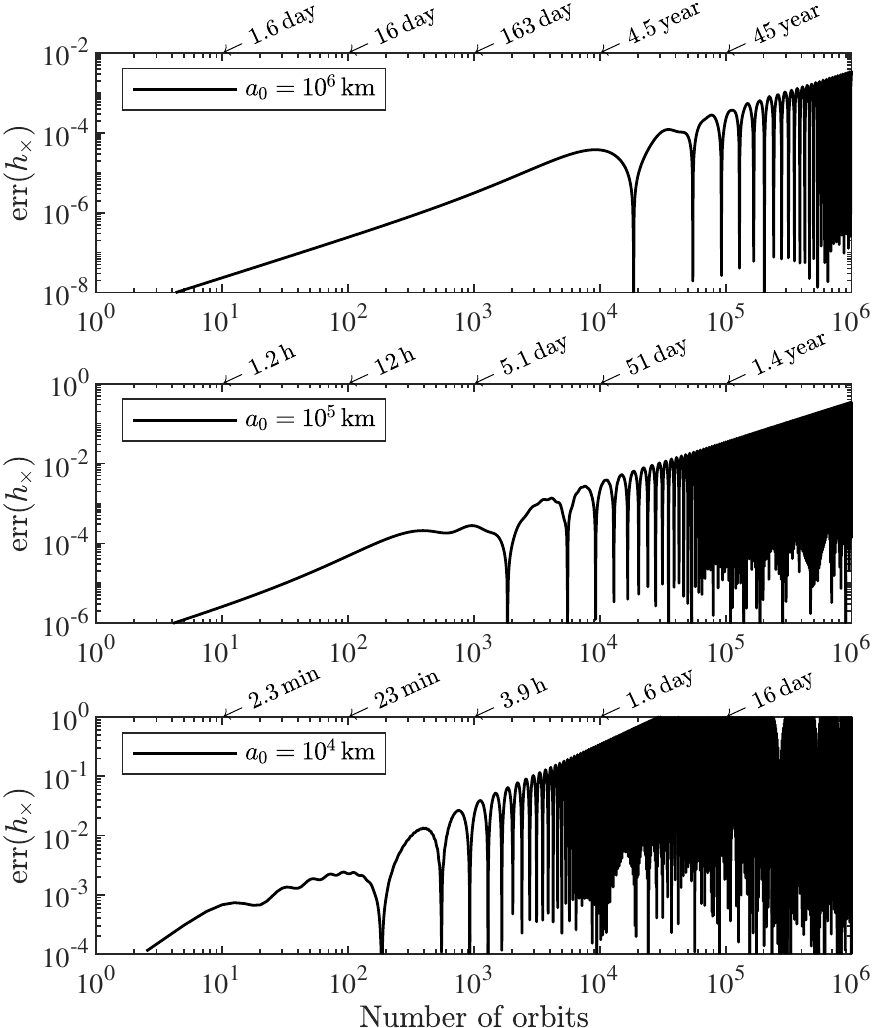}\vspace{-0.1cm}
    \caption{{\bf Left:} Relative errors caused by the fact of neglecting the dipole-dipole interaction in the computation of the mode polarization $h_+$. {\bf Right:} Similar than left panel but for the mode polarization $h_\times$.}
    \label{fig:modeGW}
\end{figure}

The evolution of $\mathrm{err}(h_+)$ and $\mathrm{err}(h_\times)$ are depicted in Fig. \ref{fig:modeGW} for different values of the initial semi-major axis. It is seen that the fact of neglecting the magnetic dipole-dipole interaction can lead to errors of the level of $100\%$ in a span of few days when the WD binary is in close orbit (i.e., for $a_0=10^4\ \mathrm{km}$).

\section{Conclusions}

In this work, we demonstrated that the dipole-dipole magnetic interaction within binary systems generates a secular drift of the argument of the periastron. The rate of the drift is proportional to the product of magnetic moments of the stars meaning that only a binary where both companions have very high magnetic moments have a chance to generate a secular drift sufficiently high for being observed by LISA. The only stars being capable of such high magnetic moments are white dwarfs even though their magnetic fields are lower than for highly magnetized NS. For a binary of highly magnetized WD in a close orbit (i.e., the LISA high frequency part) not considering the magnetic effects can generate 100\% errors in few days of observations. In this work, we neglected tidal effects for simplicity even though they could be important to account for while modeling the long term dynamics of binary systems. In addition, we assumed adiabatic magnetic fields but the more the orbit is shrinking the less the adiabatic hypothesis is accurate since internal currents might become distorted by external fields. In a future work, we aim at modeling binary systems in a coherent vision including tidal and magneto-hydrodynamics. 

\begin{acknowledgements}
  A.B. is grateful to CNES for financial support. The authors thank A.-S. Brun and A. Strugarek for fruitful discussions.
\end{acknowledgements}

\bibliographystyle{aa}  
\bibliography{bourgoin_S03} 

\end{document}